\documentclass{sigkddExp}
\usepackage{hyperref}
\usepackage[table,xcdraw]{xcolor}
\usepackage[nocompress]{cite}

\begin{document}
%
% --- Author Metadata here ---
% -- Can be completely blank or contain 'commented' information like this...
%\conferenceinfo{WOODSTOCK}{'97 El Paso, Texas USA} % If you happen to know the conference location etc.
%\CopyrightYear{2001} % Allows a non-default  copyright year  to be 'entered' - IF NEED BE.
%\crdata{0-12345-67-8/90/01}  % Allows non-default copyright data to be 'entered' - IF NEED BE.
% --- End of author Metadata ---

\title{Two-dimensional Sentiment Analysis of text}
%\subtitle{[Extended Abstract]
% You need the command \numberofauthors to handle the "boxing"
% and alignment of the authors under the title, and to add
% a section for authors number 4 through n.
%
% Up to the first three authors are aligned under the title;
% use the \alignauthor commands below to handle those names
% and affiliations. Add names, affiliations, addresses for
% additional authors as the argument to \additionalauthors;
% these will be set for you without further effort on your
% part as the last section in the body of your article BEFORE
% References or any Appendices.

\numberofauthors{1}
%
% You can go ahead and credit authors number 4+ here;
% their names will appear in a section called
% "Additional Authors" just before the Appendices
% (if there are any) or Bibliography (if there
% aren't)

% Put no more than the first THREE authors in the \author command
%%You are free to format the authors in alternate ways if you have more 
%%than three authors.

\author{
%% e-mail address with \email.
\alignauthor Rahul Tejwani \\
       \affaddr{University at Buffalo}\\
       \affaddr{Buffalo, New York}\\
       \email{rahultej@buffalo.edu}
}
\date{15 May 2014}
\maketitle
\begin{abstract}
Sentiment Analysis aims to get the underlying viewpoint of the text, which 
could be anything that holds a subjective opinion, such as  an online review,
Movie rating, Comments on Blog posts etc.  
\\*
\\
This paper presents a novel approach that classify text in two-dimensional 
Emotional space, based on
the sentiments of the author.
The approach uses existing lexical resources   
to extract feature set, which is trained 
using Supervised Learning techniques.

\end{abstract}

\section{Introduction}
With the recent growth of online reviews, social media and
blogs, there has been a lot of attention to mine for subjective 
information. These sites contains huge amount of data that has 
loads of subjective information. 
\\

Some of the challenges in Sentiment Analysis are: People
express opinions in complex ways, in opinion texts, lexical
content alone can be misleading. Humans tend to express
a lot of remarks in the form of sarcasm, irony, implication,
etc. which is very difficult to interpret. For Example- ``How
can someone sit through the movie'' is extremely negative
sentiment yet contains no negative lexographic word. Even
if a opinion word is present in the text, their can be cases
where a opinion word that is considered to be positive in one
situation may be considered negative in another situation.
People can be contradictory in their statements. Most reviews
will have both positive and negative comments.
Sometimes even other people have difficulty understanding
what someone thought based on a short piece
of text because it lacks context. A good example would be
``The laptop is good but I would prefer, the operating system
which I was using'' here context about the author's operating 
sytem is missing.
\\

There is a huge demand of sentiment analysis. Before buying
 any product its a practice now, to review its rating as rated by other
 persons who are using it. Online advice and recommendations the data reveals is 
 not the only reason behind the buzz in this area. There are other reasons from 
 company' point of view like, 
 the company wants to know ``How Successful was their last campaign or product launch'' based upon 
 the reviews of users on websites like Amazon, Yelp, etc..
 \\

\section{Previous works}
A lot of research has been done in the area over the past decade. 
Main research in the area of Sentiment Analysis and opinion mining 
are: sentiment classification, feature 
based Sentiment classification and opinion 
summarization. Sentiment classification deals with 
classifying entire documents or text or review according to the opinions 
towards certain objects. Feature-based Sentiment 
classification on the other hand considers the opinions on 
features of certain objects. For example, in reviews related to laptops
classifying the sentiments only on the basis screen quality.
\\*
\\
In one of the poineer work [2], 
the authors present a method of subjectivity 
identification for sentiment analysis based on minimum cuts. This is important 
because the irrelevant data from the reviews could be 
eliminated. 
The problem is viewed as a classification task and 
different types of Supervised learning techniques have been used 
in this field. Some of the most common ones are naive Bayes classifier, Support Vector 
Machine[13] , Maxmimum Entropy [1] etc. Even some graph based techniques [4] are also used.
\\*
\\
Languages that have been studied mostly are English 
and Chinese. Presently, there are few researches 
conducted on sentiment classification for other languages 
like Arabic, Spanish, Italian and Thai. The presented 
work focuses on English language only.

\section{Approach}
This paper uses Thayer's Model of human emotion [5], to classify text.
This two dimensional approach 
adopts the theory that human emotion
can be obtained by: 
Stress (negative polarity/positive polarity) and Energy (low intensity/high intensity), and 
divides it into four broad classes: Satisfied, 
Sad, Exuberent and Angry.
\\*
\\
Two binary classifiers were trained. First was trained to get the 
polarity (positive or nagative) of the text. While the second was trained 
on intensity (low or high) of the text. Figure 1 illustrates the approach.
\begin{figure}[h!]
  \caption{Thayer's Model}
  \centering
    \includegraphics[width=0.5\textwidth]{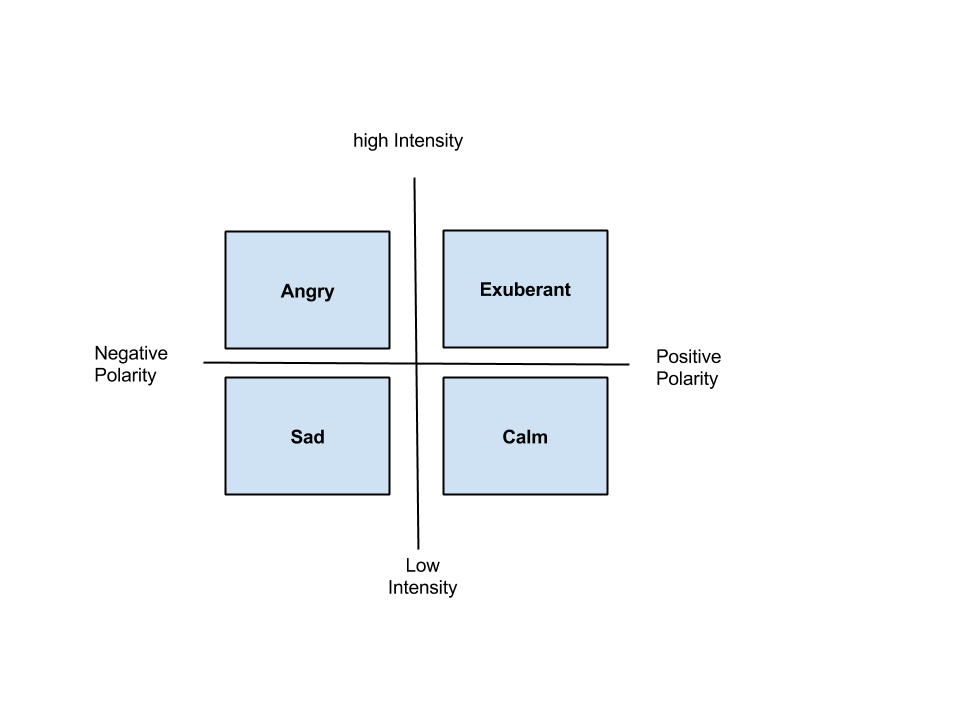}
\end{figure}

\subsection{Polarity}
Some existing lexicon resources like Sentiwordnet 3.0 [7] 
and General Inquirer [9] were used to extract some features
from the text. These features were trained using support vector 
machine, to predict the binary class label.

\subsubsection{Features}
Sentiword score of each review was used as a feature. The score
was calculated using the weighted average of the all the synsets (Synonym set, Wordnet [8])
of each word. The weights assigned were based on the ranks of
synsets as in wordnet. Thus, giving a value between -1 to 1 
depending upon the polarity. Where negative score implies negative polarity and 
vice versa while 0 being the neutral or no polarity.

\begin{equation}
 Base Score = pos - neg
\end{equation}

\begin{equation}
Polarity Score=base score + \frac{1}{2}*first + \frac{1}{3}*second..
\end{equation}

\begin{equation}
total = 1/1 + 1/2+ 1/3 +.......
\end{equation}
\begin{equation}
final score (each word)  = Polarity score/total 
\end{equation}

Let the sum of the sentiword score for each sentence be 'S' and nubmer of sentences be 'n'. 
For each sentence in the review, negation words(like: not,cant,wasnt, barely etc.)
were detected. For presence of a negation word the sentence score was multiplied by
-1. Finally each sentence score was averaged to get the Sentiword based score for 
each review.
\begin{equation}
Total Review Score  = \frac{\varSigma S}{n} 
\end{equation}

A set of 178 features were based on the frequency of 
categories marked by General Inuirer. For each word a set of
category/labels has been marked. Categories includes positive, negative,
active, passive, direct, indirect, etc.

Some other features like emoticons, number of words in quotes were also used.

\begin{table}[h]
\caption {Classification of Emoticons} \label{tab:title} 
\begin{tabular}{|p{2cm}|p{5.5cm}|}
\hline
Positive Emoticons & :) :{]} :\} :o) :o{]} :o\}:-{]} :-) :-\} =) ={]} =\}=\textasciicircum {]} =\textasciicircum ) =\textasciicircum \} :B :-D :-B :-p \\ \hline
Negative Emoticons & :-(  =(  ={[}  =\{  =\textasciicircum ;  :’(   :’{[}  :’\{  =’\{  =’(   =’{[}   =\textbackslash   :/                              \\ \hline
\end{tabular}
\end{table}

\subsection{Intensity}
To determine whether intensity of the text is high or low 
supervised learning approach has been used. Have extracted 184 
features and a support vector Model has been trained. 

\subsubsection{Features}
Some of the features used are:
all capital text, for example:
``I am EXTREMELY unhappy''. Elongated words 
have also been used as one of the features. Its a common
practice especially in online reviews that people use,
elongated words. For example ``The pizza was verryyyyyy verryyyyyy gooood!!''
Another feature includes the count of exlaimation marks, people tend to put exlamation
marks to show the level of their excitement, for example ``The coffee was too cold!!!''
Count of Adverbs was also taken. 

Finally the frequency of 178
categories from General Inquirer were also included. The frequency was
used in the same way as in Polarity.

\subsection{Dataset Creation}
There are many standard annotated Dataset available to
train polarity like, The movie review Dataset [2]. 
However no standard annotated set was available, to train 
for Intensity (low or high). 

\subsubsection{Using Yelp's Reviews to create Dataset}
The yelp's dataset [14] contains over 100,000 reviews. Each
review was marked with 1- 5 stars. Created a dataset of 5000 
(2500 positive and 2500 negative)
reviews for training polarity and another 5000 (2500 low intensity and 2500 high intensity)
to train intensity. Used the following proxy:

Considered ratings with 5 star as Positive polarity

Considered ratings with 1 star as Negative polarity.

Considered ratings with 1 or 5 star as High Intensity.

Considered ratings with 3 star as Low Intensity.

\section{Results}
The following Results are using 10 fold 
cross-validation on the dataset of size 5000. 
LIBSVM [15] is used to train a vector model.
\subsection{Results for Polarity}
The total mean accuracy achieved was 81.60\% +\//- 1.92\%

\begin{table}[h]
\caption {Result Polarity} \label{tab:title} 
\begin{tabular}{|c|c|c|l|}
\hline
\textit{}    & \textit{true ``Pos.''} & \textit{true ``Neg.''} & Class Precision \\ \hline
pred ``Pos.''       & 2143            & 563             & 79.19\%         \\ \hline
pred ``Neg.''       & 357             & 1937            & 84.44\%         \\ \hline
Class Recall & 85.72\%         & 77.48\%         &                 \\ \hline
\end{tabular}
\end{table}

\subsection{Results for Intensity}
The total mean accuracy achieved was 67.14\% +\//- 1.22\%. The details are shown
in table 3.

\begin{table}[h]
\caption {Result Intensity} \label{tab:title} 
\begin{tabular}{|c|c|c|l|}
\hline
\textit{}    & \textit{true ``Low''} & \textit{true ``high''} & Class Precision \\ \hline
pred ``Low''       & 1556            & 699             & 69.00\%         \\ \hline
pred ``High''       & 944             & 1801            & 65.61\%         \\ \hline
Class Recall & 62.24\%         & 72.04\%         &                 \\ \hline
\end{tabular}
\end{table}

\subsection{Other Approaches tried}
The following techniques were tried after removing 
stopwords and spell correction. The mean accuracy is
for 10 fold cross validation on LIBSVM.

\begin{table}[h]
\caption {Other Techniques for Polarity} \label{tab:title} 
\begin{tabular}{|c|c|}
\hline
\rowcolor[HTML]{9B9B9B} 
\textit{Techniques}                & \textit{\begin{tabular}[c]{@{}c@{}}Mean Accuracy\\ Polarity\end{tabular}} \\ \hline
All unigrams (20,000+)             & 67.1\%                                                                    \\ \hline
Adj \& Adverb with stemming (4177) & 68.6\%                                                                    \\ \hline
Adj and Adverb no Stemming         & 69.3\%                                                                    \\ \hline
Only Adjectives                    & 68.0\%                                                                    \\ \hline
Top 2000 words                     & 71.2\%                                                                    \\ \hline
\end{tabular}
\end{table}

The model did not performed very well while using
all the unigrams. One possible reason could be, that
the feature size was huge. It performed slightly better 
when only used Adjective and Adverbs. Stemming [10] the 
unigrams had almost no effect on the results. 
\\*
\\
The results were improved just by using the K-top
words occuring in the corpus, which was 2000 in this case.
One of the shortcomings with all the approaches mentioned 
in the table is that all of them are dependant on the training
Dataset. Thus, the trained model is specific to the domain and the types 
of words used in the dataset. These model will not be 
as effective for all types of text. For example, words like
Coffee, restaurent, movie, yummy, pizza etc. had high frequency in the presented
Dataset, which are not that common in a more General scenario.

\begin{center}
\begin{table}[h]
\caption {Other Techniques for Intensity} \label{tab:title} 
\begin{tabular}{|c|c|}
\hline
\rowcolor[HTML]{9B9B9B} 
\textit{Techniques}                & \textit{\begin{tabular}[c]{@{}c@{}}Mean Accuracy\\ Intensity\end{tabular}} \\ \hline
Adverb with stemming (2086)            & 58.7\%                                                                    \\ \hline
Adverb no Stemming 			& 58.4\%                                                                    \\ \hline
Frequency of 100 related categories      & 65.4\%                                                                    \\ \hline
\end{tabular}
\end{table}
\end{center}

The table above presents some of the techniques
that did not work out well. Considering unigram features, 
for a relatively small dataset did not worked out. Even omitting
some of the categories (Categories, in General Iquirer) that were generic, the
results were not good. Some of the categories that were omitted in the
above approach are, 'Doctrine', 'Economics', 'religion'
'Politics' etc.

\subsection{Mapping in 2-Dimentional emotional space}
Using Thayer's model, the following are the
Mappings in 2-Dimensional Emotional space, using 
the binary labels of Polarity and Intensity as shown in 
table 6.
\begin{center}
\begin{table}[h]
\caption {Mapping Using Thayer's Model} \label{tab:title} 
\begin{tabular}{|p{2cm}|p{2cm}|p{3cm}|}
\hline
\rowcolor[HTML]{9B9B9B} 
Polarity & Intensity & Emotion            \\ \hline
Positive & Low       & Satisfied/Calm     \\ \hline
Positive & High      & Exuberant/ Excited \\ \hline
Negative & Low       & Sad/Down           \\ \hline
Negative & High      & Angry/Agitated     \\ \hline
\end{tabular}
\end{table}
\end{center}

\section{Conclusions}
Discovered a unique way to classify text in 
two-dimensions and map to a emotion using Thayer's Model.
Did an analysis of various different techniques and compared
their results. 
\\

The proposed model for polarity was able to achieve results (81.60\%) which is 
comparable to current state of the art techniques. The approach used
lexicon based features to train the model. The Learned model does not use 
any corpus specific features for the training. The model
uses predefined set of categories that are generic. These 
catagories can be applied to any English word. Although, some
of the features used are dependant on ``Internet Lingo'', but 
they are not specific to a domain. Therefore the model could,
be applied to the text of various other domains.
\\

Used Intensity of text as a seperate dimension and created an annotated
dataset. All the features used for training Intensity were not
specific to the dataset. This approach can be used in any domain.

%ACKNOWLEDGEMENTS are optional
\section{Acknowledgements}
I would like to express my gratitude Yelp, for 
providing us the review data for research purposes.
I would specially like to thank Professor Rohini Srihari, for 
her guidance throughout the research.
I would also like to thank Puneet Singh, for all 
the brainstorming sessions and to the Authors of all 
the references list below, for their inspiring work.

\bibliographystyle{unsrt}
\bibliography{sigproc}

\begin{thebibliography}{10}

\bibitem{pang2002thumbs}
Bo~Pang, Lillian Lee, and Shivakumar Vaithyanathan.
\newblock Thumbs up?: sentiment classification using machine learning
  techniques.
\newblock In {\em Proceedings of the ACL-02 conference on Empirical methods in
  natural language processing-Volume 10}, pages 79--86. Association for
  Computational Linguistics, 2002.

\bibitem{pang2004sentimental}
Bo~Pang and Lillian Lee.
\newblock A sentimental education: Sentiment analysis using subjectivity
  summarization based on minimum cuts.
\newblock In {\em Proceedings of the 42nd annual meeting on Association for
  Computational Linguistics}, page 271. Association for Computational
  Linguistics, 2004.

\bibitem{liu2010sentiment}
Bing Liu.
\newblock Sentiment analysis and subjectivity.
\newblock {\em Handbook of natural language processing}, 2:627--666, 2010.

\bibitem{jin2008mining}
Wei Jin.
\newblock {\em Mining hidden associations in text corpora through concept chain
  and graph queries}.
\newblock ProQuest, 2008.

\bibitem{thayer1989biopsychology}
Robert~E Thayer.
\newblock {\em The biopsychology of mood and arousal}.
\newblock Oxford University Press, 1989.

\bibitem{hu2006opinion}
Minqing Hu and Bing Liu.
\newblock Opinion extraction and summarization on the web.
\newblock In {\em AAAI}, volume~7, pages 1621--1624, 2006.

\bibitem{baccianella2010sentiwordnet}
Stefano Baccianella, Andrea Esuli, and Fabrizio Sebastiani.
\newblock Sentiwordnet 3.0: An enhanced lexical resource for sentiment analysis
  and opinion mining.
\newblock In {\em LREC}, volume~10, pages 2200--2204, 2010.

\bibitem{miller1995wordnet}
George~A Miller.
\newblock Wordnet: a lexical database for english.
\newblock {\em Communications of the ACM}, 38(11):39--41, 1995.

\bibitem{stone1966general}
Philip~J Stone, Dexter~C Dunphy, and Marshall~S Smith.
\newblock The general inquirer: A computer approach to content analysis.
\newblock 1966.

\bibitem{porter1980algorithm}
Martin~F Porter.
\newblock An algorithm for suffix stripping.
\newblock {\em Program: electronic library and information systems},
  14(3):130--137, 1980.

\bibitem{mullen2004sentiment}
Tony Mullen and Nigel Collier.
\newblock Sentiment analysis using support vector machines with diverse
  information sources.
\newblock In {\em EMNLP}, volume~4, pages 412--418, 2004.

\bibitem{tejwani2014sentiment}
Rahul Tejwani.
\newblock Sentiment analysis: A survey.
\newblock {\em arXiv preprint arXiv:1405.2584}, 2014.

\bibitem{singh2010architecture}
Puneet Singh, Ashutosh Kapoor, Vishal Kaushik, and Hima~Bindu Maringanti.
\newblock Architecture for automated tagging and clustering of song files
  according to mood.
\newblock {\em International Journal of Computer Science Issues (IJCSI)}, 7(4),
  2010.

\end{thebibliography}

\nocite{*}

% sigproc.bib is the name of the Bibliography in this case
% ACM needs 'a single self-contained file'!
% SIGKDD: balancing columns messes up the footers: Sunita Sarawagi, Jan 2000.
% \balancecolumns

[14] Yelp's Review Dataset: 

\url{https://www.yelp.com/academic\textunderscore dataset}
\\

[15] LIBSVM -- A Library for Support Vector Machines: 

\url{http://www.csie.ntu.edu.tw/~cjlin/libsvm/}

% That's all folks!
\end{document}